\documentclass[thmsa,letterpaper]{article}
\usepackage{sw20lart}

\input{tcilatex}
\begin{document}

\author{S. Manoff \\
\textit{Bulgarian Academy of Sciences}\\
\textit{Institute for Nuclear Research and Nuclear Energy}\\
\textit{Department of Theoretical Physics}\\
\textit{Blvd. Tzarigradsko Chaussee 72}\\
\textit{1784 - Sofia, Bulgaria}}
\date{E-mail address: smanov@inrne.bas.bg\\
Published in\textbf{: Perspectives in Complex Analysis, Differential
Geometry and Mathematical Physics. World Scientific, Singapore, 2001, 190-200%
}}
\title{\textsc{Lagrangian fluid mechanics}}
\maketitle

\begin{abstract}
The method of Lagrangians with covariant derivative (MLCD) is applied to a
special type of Lagrangian density depending on scalar and vector fields as
well as on their first covariant derivatives. The corresponding
Euler-Lagrange's equations and energy-momentum tensors are found on the
basis of the covariant Noether's identities.
\end{abstract}

\section{Introduction}

In the recent years it has been shown that every classical field theory \cite
{Manoff-1}, \cite{Manoff-2} could be considered as a theory of a continuous
media with its kinematic characteristics. On the other side, ''perfect
fluids are equivalent to elastic media when the later are homogeneous and
isotropic'' \cite{Brown}. The theory of fluids is usually based on a state
equation and on a variational principle \cite{Taub}, \cite{Schutz-1}, \cite
{Schutz-2} with a given Lagrangian density depending on the variables
constructing the equation of state (rest mass density, entropy, enthalpy,
temperature, pressure etc.) and the 4-velocity of the points in the fluid.
There are two canonical types of description of a fluid: the Eulerian
picture and the Lagrangian picture. In the Eulerian picture, a point of the
fluid and its velocity are identified with the state and the velocity of an
observer moving with it. This means that every point in the fluid has a
velocity which coincides with the velocity of an observer moving with it. In
the Lagrangian picture, the observer has different velocity from the
velocity of a point of the fluid observed by him. He observes the
projections of the velocities of the points of the fluids from its own frame
of reference. This means that an observer does not move with the points of
the fluid but measures their velocities projecting them on its own velocity
and standpoint. In other words, the Eulerian picture is related to the
description of a fluid by an observer moving locally with its points, and
the Lagrangian picture is related to the description of a fluid by an
observer not moving locally with its points. In the Eulerian picture the
velocity of the points (the velocity of an observer) in a fluid appears as a
field variable in the Lagrangian density describing the fluid. In the
Lagrangian picture the velocity of an observer could be considered as an
element of the projection formalism [called $(n+1)-1$ - formalism, $\dim M=n$%
, where $M$ is a differential manifold with dimension $n$, considered as a
model of the space ($n=3$) or the space-time ($n=4$)]. The Lagrangian
invariant $L$ by the use of which a Lagrangian density $\mathbf{L}$ is
defined as $\mathbf{L}=\sqrt{-d_g}\cdot L$, where $d_g=\det (g_{ij})$. The
tensor $g=g_{ij}\cdot dx^i.dx^j=\frac 12\cdot g_{ij}.(dx^i\otimes
dx^j+dx^j\otimes dx^i)$ is the metric tensor field in $M$ determining the
possibility for introducing the square of the line element $ds^2=g(d,d)=g_{%
\overline{i}\overline{j}}\cdot dx^i\cdot dx^j$ with $d=dx^i\cdot \partial
_i\in T(M):=\cup _xT_x(M)$ [$T_x(M)$ is the tangent space to the point $x\in
M$, $:=$ means by definition, $i,j,k,...=1$, $2$, ... $n$], interpreted as
the infinitesimal distance between two points: point $x$ with co-ordinates $%
x^i$ and point $\overline{x}$ with co-ordinates $x^i+dx^i$. The
differentiable manifold $M$ is considered here as a space with contravariant
and covariant affine connections (whose components differ not only by sign)
and metrics, i. e. $M\equiv (\overline{L}_n,g)$-space \cite{Manoff-3}, \cite
{Manoff-4}. In $(\overline{L}_n,g)$-spaces with $n=4$ the Lagrangian
invariant $L$ could be interpreted as the pressure $p$ in a material system.
This interpretation is based on the fact that for $n=4$, the invariant $L$
has the same dimension as the pressure $p:[p]=\frac{[\text{force}]}{[\text{%
2-surface}]}=[L]$ of a physical system, described by its Lagrangian density $%
\mathbf{L}$, i.e. $\mathbf{L}=\sqrt{-d_g}\cdot p$. It allows us to consider $%
L$ as the pressure of a fluid. On this basis, we could formulate the
following hypothesis:

\textit{Hypothesis}. The pressure $p$ of a dynamical system could be
identified with the Lagrangian invariant $L$ of this system, i. e. 
\begin{equation}
p:=L(g_{ij},\,\,\,\,g_{ij;k},\,\,\,\,\,g_{ij;k;l},\,\,\,\,V^A\,_B,\,V^A%
\,_{B;i},\,\,\,V^A\,_{B;i;j})\text{ ,}  \label{1.1}
\end{equation}

\noindent where $g_{ij}$ are the components of the metric tensor field of
the space (or space-time), where the system exists and $V^A\,_B$ are
components of tensor fields describing the state of this system. $g_{ij;k}$, 
$g_{ij;k;l}$ are the first and second covariant derivative of the metric
tensor $g$ with respect to a covariant affine connection $P$ and $%
V^A\,_{B;i} $, $V^A\,_{B;i;j}$ are the first and second covariant
derivatives of the tensor fields $V\in \otimes ^k\,_l(M)$ with respect to a
contravariant affine connection $\Gamma $ and a covariant affine connection $%
P$.

The hypothesis could be considered from two different points of view:

1. If the pressure in a dynamical system is given, i.e. if a state equation
of the type 
\[
p:=p(g_{ij},\,\,\,\,g_{ij;k},\,\,\,\,\,g_{ij;k;l},\,\,\,\,V^A\,_B,\,V^A%
\,_{B;i},\,\,\,V^A\,_{B;i;j}) 
\]

\noindent is given, then $p$ could be identified with the Lagrangian
invariant $L$ of the system. On this basis a Lagrangian theory of a system
with a given pressure $p$ could be worked out.

2. If a Lagrangian invariant $L$ of a Lagrangian density $\mathbf{L}=\sqrt{%
-d_g}\cdot L$ is given, i.e. if 
\[
L:=L(g_{ij},\,\,\,\,g_{ij;k},\,\,\,\,\,g_{ij;k;l},\,\,\,\,V^A\,_B,\,V^A%
\,_{B;i},\,\,\,V^A\,_{B;i;j}) 
\]

\noindent is given, then $L$ could be identified with the pressure $p$ of
the dynamical system. On these grounds a Lagrangian theory of fluids could
be worked out with a given pressure $p$. In this case $L:=p$.

Therefore, we can distinguish two cases:

Case 1. $L:=p$ with given $p\in C^r(M)$.

Case 2. $p:=L$ with given $L\in C^r(M)$.

In the present paper we consider a Lagrangian fluid mechanics by the use of
the method of Lagrangians with covariant derivatives of the type (\ref{1.1}%
). The Lagrangian invariant $L:=p$ (considered as the pressure $p$) could
depend on the velocity vector field $u\in T(M)$ of the points in the fluid,
on other vector fields $\xi \in T(M)$ orthogonal (or not orthogonal) to $u$ [%
$g(u,\xi )=0$ or $g(u,\xi )\neq 0$], on scalar fields $f_N=f_N(x^k)$ ($N=1$, 
$2$, ..., $m\in \mathbf{R}$) describing the state of the fluid as well as on
the (first) covariant derivatives of the vector fields $u$ and $\xi $. For a
given in its explicit form Lagrangian density $\mathbf{L}$ and its
Lagrangian invariant $L$ respectively, we find the covariant
Euler-Lagrange's equations and the corresponding energy-momentum tensors.

\section{Lagrangian density and Lagrangian invariant}

Let us consider a Lagrangian density of the type $\mathbf{L}=\sqrt{-d_g}%
\cdot L$ with Lagrangian invariant $L$ in the form: 
\begin{eqnarray}
L &:&=p:=p_0+a_0\cdot \rho \cdot e+b_0\cdot g(\nabla _u(\rho \cdot
u),u)+c_0\cdot g(\nabla _u(\rho \cdot \xi ),\xi )  \label{1.2} \\
&&+f_0\cdot g(\nabla _u(\rho \cdot \xi ),u)+h_0\cdot g(\nabla _u(\rho \cdot
u),\xi )  \nonumber \\
&&+\kappa _0\cdot \rho \cdot \frac{m_0}{\sqrt{g(\xi ,\xi )}}+k_0\cdot \frac{%
M_0\cdot \rho }{[g(u,u)]^k\cdot [g(\xi ,\xi )]^m}  \nonumber \\
&&+k_1\cdot \rho ^r+p_1(f_N,\,\,\,\,\,f_{N,i},\text{ }u^i)\,\,\,\,\,\text{.}
\nonumber
\end{eqnarray}

The quantities $p_0$, $a_0$, $b_0$, $c_0$, $f_0$, $h_0$, $\kappa _0$, $k_0$, 
$m_0$, $M_0$, and $k_1$ are constants, $k$, $m$, $r$ are real numbers, $%
f_N:=f_N(x^k)\in C^r(M)$ are real functions identified as thermodynamical
and kinematical variables, $N\in \mathbf{N}$. The function $\rho =\rho (x^k)$
is an invariant function with respect to the co-ordinates in $M$. The vector
field $u=u^i\cdot \partial _i\in T(M)$ is a contravariant non-isotropic
(non-null) vector field with $g(u,u):=e:\neq 0$. The vector field $\xi \in
T(M)$ is a contravariant vector field with $g(\xi ,\xi ):\neq 0$ in the
cases when $\kappa _0\neq 0$, $m_0\neq 0$, $k_0\neq 0$, $M_0\neq 0$, $\rho
(x^k)\neq 0$ and $g(\xi ,\xi )=0$ or $g(\xi ,\xi )=0$ if $\kappa
_0=m_0=k_0=M_0=0$. The constants $a_0$, $b_0$, $c_0$, $f_0$, $h_0$, $\kappa
_0$, $k_0$, and $k_1$ could also be considered as Lagrangian multipliers to
the corresponding constraints of 1. kind: 
\begin{eqnarray}
a_0 &:&\rho \cdot e=0\text{ ,}  \label{1.3} \\
b_0 &:&g(\nabla _u(\rho \cdot u),u)=0\text{ ,}  \nonumber \\
c_0 &:&g(\nabla _u(\rho \cdot \xi ),\xi )=0\text{ ,}  \nonumber \\
f_0 &:&g(\nabla _u(\rho \cdot \xi ),u)=0\text{ ,}  \nonumber \\
h_0 &:&g(\nabla _u(\rho \cdot u),\xi )=0\text{ ,}  \nonumber \\
\kappa _0 &:&\rho \cdot \frac{m_0}{\sqrt{g(\xi ,\xi )}}=0\text{ ,}  \nonumber
\\
k_0 &:&\frac{M_0\cdot \rho }{[g(u,u)]^k\cdot [g(\xi ,\xi )]^m}=0\text{ ,} 
\nonumber \\
k_1 &:&\rho ^r=0\rightarrow \rho =0\text{.}  \nonumber
\end{eqnarray}

Depending on the considered case the corresponding constants could be chosen
to be or not to be equal to zero.

\subsection{Representation of the Lagrangian invariant $L$ in a useful for
variations form}

For finding out the Euler-Lagrange's equations one needs to represent the
Lagrangian invariant $L$ in a form, suitable for the application of the
method of Lagrangians with covariant derivatives \cite{Manoff-5}. For this
reason the pressure $p=L$ could be written in the form 
\begin{equation}
p=p_0+p_1(f_N\text{, }uf_N)+k_1\cdot \rho ^r+\rho \cdot f+(u\rho )\cdot b%
\text{ ,}  \label{1.4}
\end{equation}

\noindent where 
\begin{eqnarray}
f &:&=a_0\cdot e+b_0\cdot g(a,u)+c_0\cdot g(\nabla _u\xi ,\xi )+f_0\cdot
g(\nabla _u\xi ,u)+h_0\cdot g(a,\xi )  \nonumber \\
&&+\kappa _0\cdot \frac{m_0}{\sqrt{g(\xi ,\xi )}}+k_0\cdot \frac{M_0}{%
[g(u,u)]^k\cdot [g(\xi ,\xi )]^m}\text{ ,}  \label{1.5} \\
b &:&=b_0\cdot e+c_0\cdot g(\xi ,\xi )+(f_0+h_0)\cdot l\text{ ,}  \nonumber
\\
a &:&=\nabla _uu=u^i\,_{;j}\cdot u^j=a^i\cdot \partial _i\text{ , \thinspace
\thinspace \thinspace \thinspace \thinspace \thinspace \thinspace }%
l:=g(u,\xi )\text{ .}  \nonumber
\end{eqnarray}

In a co-ordinate basis $f$, $b$, $a$, and $l$ have the form: 
\begin{eqnarray}
f &=&g_{\overline{k}\overline{l}}\cdot [a_0\cdot u^k\cdot u^l+b_0\cdot
u^k\,_{;m}\cdot u^m\cdot u^l+c_0\cdot \xi ^k\,_{;m}\cdot u^m\cdot \xi ^l 
\nonumber \\
&&+f_0\cdot \xi ^k\,_{;m}\cdot u^m\cdot u^l+h_0\cdot u^k\,_{;m}\cdot
u^m\cdot \xi ^l]  \nonumber \\
&&+\kappa _0\cdot \frac{m_0}{\sqrt{g_{\overline{k}\overline{l}}\cdot \xi
^k\cdot \xi ^l}}+k_0\cdot \frac{M_0}{[g_{\overline{k}\overline{l}\cdot
}u^k\cdot u^l]^k\cdot [g_{\overline{m}\overline{n}}\cdot \xi ^m\cdot \xi
^n]^m}\text{ ,}  \label{1.6} \\
b &=&g_{\overline{k}\overline{l}}\cdot [b_0\cdot u^k\cdot u^l+c_0\cdot \xi
^k\cdot \xi ^l+(f_0+h_0)\cdot u^k\cdot \xi ^l]\text{ \thinspace \thinspace
\thinspace .}  \nonumber
\end{eqnarray}

Therefore, we can consider $p$, $f$, and $b$ as functions of the field
variables $f_N$, $\rho $, $u$, $\xi $, and $g$ as well as of their
corresponding first covariant derivatives.

\section{Euler-Lagrange's equations for the variables on which the pressure $%
p$ depends}

We can apply now the method of Lagrangians with covariant derivatives to the
explicit form of the pressure $p$ and find the Euler-Lagrange's equations
for the variables $f_N$, $\rho $, $u$, $\xi $, and $g$. After long (but not
so complicated computations) the Euler-Lagrange's equations follow in the
form:

1. Euler-Lagrange's equations for the thermodynamical functions $f_N$: 
\begin{equation}
\frac{\partial p_1}{\partial f_N}-(\frac{\partial p_1}{\partial f_{N,i}}%
)_{;i}+q_i\cdot \frac{\partial p_1}{\partial f_{N,i}}=0\text{ ,}  \label{2.1}
\end{equation}

\noindent where 
\begin{eqnarray}
q_i &=&T_{ki}\,^k-\frac 12\cdot g^{\overline{k}\overline{l}}\cdot
g_{kl;i}+g_k^l\cdot g_{l;i}^k\text{ ,}  \label{2.2} \\
T_{ki}\,^k &=&g_l^k\cdot T_{ki}\,^l\text{ , \thinspace \thinspace \thinspace
\thinspace \thinspace \thinspace \thinspace \thinspace }T_{ki}\,^l=\Gamma
_{ik}^l-\Gamma _{ki}^l\text{ ,\thinspace \thinspace \thinspace \thinspace
\thinspace \thinspace \thinspace \thinspace \thinspace \thinspace \thinspace 
}g_{l;i}^k=\Gamma _{li}^k+P_{li}^k\text{ \thinspace \thinspace \thinspace .}
\nonumber
\end{eqnarray}

2. Euler-Lagrange's equation for the function $\rho $: 
\begin{equation}
ub=k_1\cdot r\cdot \rho ^{r-1}+f+(q-\delta u)\cdot b\,\,\,\,\,\text{,}
\label{2.3}
\end{equation}

\noindent where 
\begin{equation}
ub=u^k\cdot b_{,k}\text{ ,\thinspace \thinspace \thinspace \thinspace
\thinspace \thinspace \thinspace \thinspace \thinspace \thinspace \thinspace 
}q=q_i\cdot u^i\text{ \thinspace \thinspace \thinspace \thinspace \thinspace
\thinspace , \thinspace \thinspace \thinspace \thinspace \thinspace
\thinspace }\delta u=u^i\,_{;i}=u^i\,_{;k}\cdot g_i^k\text{ \thinspace
\thinspace \thinspace \thinspace \thinspace .}  \label{2.4}
\end{equation}

3. Euler-Lagrange's equations for the contravariant vector field $u$: 
\begin{eqnarray}
(h_0-f_0)\cdot \xi ^i\,_{;k}\cdot u^k &=&g^{ij}\cdot [b\cdot (\log \rho
)_{,j}+\frac 1\rho \cdot \frac{\partial p_1}{\partial u^j}]  \nonumber \\
&&+\{2\cdot (a_0-k_0\cdot k\cdot \frac{M_0}{[g(u,u)]^{k+1}\cdot [g(\xi ,\xi
)]^m})  \nonumber \\
&&+b_0\cdot [q-\delta u+u(\log \rho )]\}\cdot u^i  \nonumber \\
&&+\{h_0\cdot (q-\delta u)+f_0\cdot [u(\log \rho )]\}\cdot \xi ^i
\label{2.5} \\
&&+g_{\overline{k}\overline{l}}\cdot [(b_0\cdot u^l+h_0\cdot \xi ^l)\cdot
(u^k\,_{;j}-g_{j;m}^k\cdot u^m)\cdot g^{ji}  \nonumber \\
&&+(c_0\cdot \xi ^l+f_0\cdot u^l)\cdot \xi ^k\,_{;j}\cdot g^{ji}]  \nonumber
\\
&&-g^{ij}\cdot g_{\overline{j}\overline{k};m}\cdot u^m\cdot (b_0\cdot
u^k+h_0\cdot \xi ^k)\text{ \thinspace \thinspace \thinspace \thinspace
\thinspace \thinspace ,}  \nonumber
\end{eqnarray}

\noindent where 
\begin{equation}
g_{\overline{j}\overline{k};m}:=f^n\,_j\cdot f^l\,_k\cdot g_{nl;m}\text{
,\thinspace \thinspace \thinspace \thinspace \thinspace \thinspace
\thinspace \thinspace \thinspace \thinspace \thinspace }u(\log \rho
)=u^i\cdot (\log \rho )_{,i}\text{ \thinspace \thinspace \thinspace
\thinspace .}  \label{2.6}
\end{equation}

4. Euler-Lagrange's equations for the contravariant vector field $\xi $: 
\begin{eqnarray}
(f_0-h_0)\cdot a^i &=&[f_0\cdot (q-\delta u)+h_0\cdot u(\log \rho )]\cdot u^i
\nonumber \\
&&+\{c_0\cdot [q-\delta u+u(\log \rho )]  \nonumber \\
&&-(\kappa _0\cdot \frac{m_0}{[g(\xi ,\xi )]^{3/2}}+2\cdot k_0\cdot m\cdot 
\frac{M_0}{[g(u,u)]^k\cdot [g(\xi ,\xi )]^{m+1}})\}\cdot \xi ^i  \nonumber \\
&&-g^{ij}\cdot g_{\overline{j}\overline{k};m}\cdot u^m\cdot (c_0\cdot \xi
^k+f_0\cdot u^k)  \label{2.7} \\
&&-g^{ij}\cdot g_{j;m}^k\cdot u^m\cdot g_{\overline{k}\overline{l}}\cdot
(c_0\cdot \xi ^l+f_0\cdot u^l)\,\,\,\,\,\,\,\,\text{.}  \nonumber
\end{eqnarray}

5. Euler-Lagrange's equations for the covariant metric tensor $g$: 
\begin{eqnarray}
g^{ij} &=&-\frac 2p\cdot \{[b_0\cdot (u\rho )+\rho \cdot (a_0-k_0\cdot
k\cdot \frac{M_0}{[g(u,u)]^{k+1}\cdot [g(\xi ,\xi )]^m})]\cdot u^i\cdot u^j 
\nonumber \\
&&+[c_0\cdot (u\rho )-\rho \cdot (\frac{\kappa _0}2\cdot \frac{m_0}{[g(\xi
,\xi )]^{3/2}}+k_0\cdot m\cdot \frac{M_0}{[g(u,u)]^k\cdot [g(\xi ,\xi
)]^{m+1}})]\xi ^i\cdot \xi ^j  \nonumber \\
&&+\frac 12\cdot \rho \cdot [b_0\cdot (a^i\cdot u^j+a^j\cdot b^i)+h_0\cdot
(a^i\cdot \xi ^j+a^j\cdot \xi ^i)  \label{2.8} \\
&&+c_0\cdot (\xi ^i\,_{;m}\cdot \xi ^j+\xi ^j\,_{;m}\cdot \xi ^i)\cdot
u^m+f_0\cdot (\xi ^i\,_{;m}\cdot u^j+\xi ^j\,_{;m}\cdot u^i)\cdot u^m] 
\nonumber \\
&&+\frac 12\cdot (u\rho )\cdot (f_0+h_0)\cdot (u^i\cdot \xi ^j+u^j\cdot \xi
^i)\}  \nonumber \\
&=&-\frac 2p\cdot [A\cdot u^i\cdot u^j+B\cdot \xi ^i\cdot \xi ^j+\frac
12\cdot \rho \cdot C^{ij}+\frac 12\cdot (u\rho )\cdot D^{ij}\text{
\thinspace \thinspace \thinspace \thinspace \thinspace \thinspace \thinspace
,}  \nonumber
\end{eqnarray}

\noindent where 
\begin{eqnarray}
A &=&b_0\cdot (u\rho )+\rho \cdot (a_0-k_0\cdot k\cdot \frac{M_0}{%
[g(u,u)]^{k+1}\cdot [g(\xi ,\xi )]^m})\text{ \thinspace \thinspace
\thinspace \thinspace \thinspace ,}  \nonumber \\
B &=&c_0\cdot (u\rho )-\rho \cdot (\frac{\kappa _0}2\cdot \frac{m_0}{[g(\xi
,\xi )]^{3/2}}+k_0\cdot m\cdot \frac{M_0}{[g(u,u)]^k\cdot [g(\xi ,\xi
)]^{m+1}})\text{ \thinspace \thinspace \thinspace \thinspace \thinspace ,} 
\nonumber \\
C^{ij} &=&b_0\cdot (a^i\cdot u^j+a^j\cdot b^i)+h_0\cdot (a^i\cdot \xi
^j+a^j\cdot \xi ^i)  \label{2.9} \\
&&+c_0\cdot (\xi ^i\,_{;m}\cdot \xi ^j+\xi ^j\,_{;m}\cdot \xi ^i)\cdot
u^m+f_0\cdot (\xi ^i\,_{;m}\cdot u^j+\xi ^j\,_{;m}\cdot u^i)\cdot u^m\text{
\thinspace \thinspace \thinspace \thinspace \thinspace \thinspace ,} 
\nonumber \\
D^{ij} &=&(f_0+h_0)\cdot (u^i\cdot \xi ^j+u^j\cdot \xi ^i)\text{\thinspace
\thinspace \thinspace \thinspace \thinspace \thinspace \thinspace \thinspace
.}  \nonumber
\end{eqnarray}

The Euler-Lagrange's equations for the different variables are worth to be
investigated in details in general as well as for every special case with a
subset of constant different from zero.

\subsection{Conditions for the pressure $p$ which follow from the
Euler-Lagrange's equations for the covariant metric field $g$}

The Euler-Lagrange's equations (ELEs) for the metric field $g$ lay down
conditions to the form of the pressure $p$ and its dependence on the other
variables. The Euler-Lagrange's equations for $g$ could be written in the
general form as 
\begin{equation}
\frac{\partial p}{\partial g_{ij}}+\frac 12\cdot g^{\overline{i}\overline{j}%
}=0\text{ \thinspace \thinspace \thinspace \thinspace .}  \label{2.10}
\end{equation}

After contracting with $g_{ij}$ and summarizing over $i$ and $j$ we obtain
the condition 
\begin{equation}
\frac{\partial p}{\partial g_{ij}}\cdot g_{ij}+\frac n2\cdot p=0\text{
\thinspace \thinspace \thinspace \thinspace \thinspace \thinspace \thinspace
\thinspace \thinspace \thinspace \thinspace \thinspace \thinspace \thinspace
\thinspace \thinspace \thinspace \thinspace }p=-\frac 2n\cdot \frac{\partial
p}{\partial g_{ij}}\cdot g_{ij}\text{ \thinspace \thinspace \thinspace
\thinspace \thinspace \thinspace \thinspace \thinspace .}  \label{2.11}
\end{equation}

On the other side, if we contract the ELEs for $p$ with $\xi _j=g_{j%
\overline{m}}\cdot \xi ^m$ or with $u_j=g_{j\overline{n}}\cdot u^n$ we
obtain the following relations respectively: 
\begin{equation}
(A^i\,_k-g_k^i)\cdot \xi ^k=0\text{ ,\thinspace \thinspace \thinspace
\thinspace \thinspace \thinspace \thinspace \thinspace \thinspace \thinspace
\thinspace \thinspace \thinspace \thinspace \thinspace \thinspace \thinspace
\thinspace \thinspace \thinspace }(A_k^i-g_k^i)\cdot u^k=0\text{ \thinspace
\thinspace \thinspace \thinspace \thinspace \thinspace \thinspace \thinspace
,}  \label{2.12}
\end{equation}

\noindent where 
\begin{equation}
A^i\,_k=-\frac 2p\cdot \frac{\partial p}{\partial g_{lj}}\cdot g_{jm}\cdot
f^m\,_k\cdot f_l\,^i\text{ \thinspace \thinspace \thinspace \thinspace .}
\label{2.13}
\end{equation}

$f^m\,_k$ are components of the contraction tensor $Sn$ \cite{Manoff-1} and $%
f^m\,_k\cdot f_m\,^i=g_k^i$. Since 
\begin{equation}
-\frac 2p\cdot \frac{\partial p}{\partial g_{ij}}\cdot g_{jk}=g_k^i\text{ ,}
\label{2.14}
\end{equation}

\noindent it follows that $A^i\,_k=g_k^i$ and therefore, the relations for $%
\xi ^i\,$and $u^i$ are identically fulfilled. The only condition remaining
for $p$ follows in the form 
\begin{eqnarray}
p &=&\frac 2{n+2}\cdot \{p_0+p_1+k_1\cdot \rho ^r  \label{2.15} \\
&&+\rho \cdot [\frac 32\cdot \kappa _0\cdot \frac{m_0}{\sqrt{g(\xi ,\xi )}}%
+(k+m+1)\cdot k_0\cdot \frac{M_0}{[g(u,u)]^k\cdot [g(\xi ,\xi )]^m}]\}\text{
\thinspace \thinspace \thinspace \thinspace .}  \nonumber
\end{eqnarray}

In the special case, when $n=4$, we have 
\begin{eqnarray}
p &=&\frac 13\cdot (p_0+p_1+k_1\cdot \rho ^r)  \nonumber \\
&&+\rho \cdot [\frac 12\cdot \kappa _0\cdot \frac{m_0}{\sqrt{g(\xi ,\xi )}}%
+\frac 13\cdot (k+m+1)\cdot k_0\cdot \frac{M_0}{[g(u,u)]^k\cdot [g(\xi ,\xi
)]^m}]\text{ \thinspace \thinspace \thinspace \thinspace \thinspace
\thinspace \thinspace \thinspace \thinspace \thinspace .}  \label{2.16}
\end{eqnarray}

By the use of the method of Lagrangians with covariant derivatives we can
also find the corresponding energy-momentum tensors.

\section{Energy-momentum tensors for a fluid with pressure $p$}

The energy-momentum tensors for the given Lagrangian density $\mathbf{L}$
could be found by the use of the method of Lagrangians with covariant
derivatives on the basis of the covariant Noether's identities \cite
{Manoff-5} 
\begin{eqnarray}
\overline{F}_i+\overline{\theta }_i\,^j\,_{;j} &\equiv &0\text{ \thinspace
\thinspace \thinspace \thinspace \thinspace \thinspace \thinspace \thinspace
\thinspace first Noether's identity,}  \label{3.1} \\
\overline{\theta }_i\,^j-\,_sT_i\,^j &\equiv &\overline{Q}_i\,^j\text{
\thinspace \thinspace \thinspace \thinspace second Noether's identity.} 
\nonumber
\end{eqnarray}

One has to distinguish three types of energy-momentum tensors: (a)
generalized canonical energy-momentum tensor $\overline{\theta }_i\,^j$; (b)
symmetric energy-momentum tensor of Belinfante $_sT_i\,^j$, and (c)
variational energy-momentum tensor of Euler-Lagrange $\overline{Q}_i\,^j$.
All three energy-momentum tensors obey the second Noether's identity. After
long computations we can find the energy-momentum tensors.

\subsection{Generalized canonical energy-momentum tensor}

The generalized energy-momentum tensor $\overline{\theta }_i\,^j$ could be
obtained in the form 
\begin{eqnarray}
\overline{\theta }_i\,^j &=&\frac{\partial p_1}{\partial f_{N,j}}\cdot
f_{N,i}+b\cdot \rho _{,i}\cdot u^j  \nonumber \\
&&+\rho \cdot g_{\overline{l}\overline{n}}\cdot [(b_0\cdot u^n+h_0\cdot \xi
^n)\cdot u^l\,_{;i}\cdot u^j+(c_0\cdot \xi ^n+f_0\cdot u^n)\cdot \xi
^l\,_{;i}\cdot u^j]  \nonumber \\
&&+g_{\overline{i}\overline{k}}\cdot \{(f_0-h_0)\cdot (\xi \rho +\rho \cdot
\delta \xi )\cdot u^j\cdot u^k-(f_0-h_0)\cdot (u\rho +\rho \cdot \delta
u)\cdot \xi ^j\cdot u^k  \nonumber \\
&&+\rho \cdot (h_0-f_0)\cdot [(\xi ^j\cdot a^k+b^j\cdot u^k)-(u^j\cdot
d^k+u^k\cdot d^j)]  \label{3.2} \\
&&+\rho \cdot (g_{\overline{l}\overline{n};m}+g_{\overline{r}\overline{n}%
}\cdot g_{l;m}^r)\cdot [(b_0\cdot u^n+h_0\cdot \xi ^n)\cdot (g^{jl}\cdot
u^k\cdot u^m-g^{lm}\cdot u^j\cdot u^k)  \nonumber \\
&&+\frac 12\cdot (c_0\cdot \xi ^n+f_0\cdot u^n)\cdot [g^{kl}\cdot (u^j\cdot
\xi ^m-u^m\cdot \xi ^j)  \nonumber \\
&&+g^{jl}\cdot (u^k\cdot \xi ^m+u^m\cdot \xi ^k)-g^{lm}\cdot (u^j\cdot \xi
^k+u^k\cdot \xi ^j)]]\}-p.g_i^j\text{ \thinspace \thinspace \thinspace
\thinspace ,}  \nonumber
\end{eqnarray}

\noindent where 
\begin{eqnarray}
a^k &=&u^k\,_{;l}\cdot u^l\text{ \thinspace ,\thinspace \thinspace
\thinspace \thinspace \thinspace \thinspace \thinspace \thinspace \thinspace
\thinspace \thinspace \thinspace }\delta u=u^m\,_{;m}\text{ ,\thinspace
\thinspace \thinspace \thinspace \thinspace \thinspace \thinspace }\delta
\xi =\xi ^l\,_{;l}\text{ ,\thinspace \thinspace \thinspace \thinspace
\thinspace \thinspace \thinspace \thinspace \thinspace \thinspace \thinspace 
}\xi \rho =\rho _{,j}\cdot \xi ^j\text{ ,\thinspace \thinspace \thinspace
\thinspace \thinspace \thinspace \thinspace \thinspace }  \nonumber \\
\text{\thinspace }b^j &=&\xi ^j\,_{;m}\cdot u^m\text{ , \thinspace
\thinspace \thinspace \thinspace \thinspace \thinspace \thinspace \thinspace 
}d^k=u^k\,_{;l}\cdot \xi ^l\text{ , \thinspace \thinspace \thinspace
\thinspace \thinspace \thinspace }f^k=\xi ^k\,_{;m}\cdot \xi ^m\text{
\thinspace \thinspace \thinspace .}  \label{3.2a}
\end{eqnarray}

\subsection{Symmetric energy-momentum tensor of Belinfante}

The symmetric energy-momentum tensor of Belinfante $_sT_i\,^j$ could be
obtained in the form 
\begin{eqnarray}
_sT_i\,^j &=&g_{\overline{i}\overline{k}}\cdot \{[b_0\cdot (u\rho +\rho
\cdot \delta u)+(f_0-h_0)\cdot (\xi \rho +\rho \cdot \delta \xi )]\cdot
u^j\cdot u^k  \nonumber \\
&&+c_0\cdot (u\rho +\rho \cdot \delta u)\cdot \xi ^j\cdot \xi ^k+h_0\cdot
(u\rho +\rho \cdot \delta u)\cdot (u^j\cdot \xi ^k+u^k\cdot \xi ^j) 
\nonumber \\
&&+\rho \cdot [b_0\cdot (u^j\cdot a^k+u^k\cdot a^j)+c_0\cdot (\xi ^j\cdot
b^k+\xi ^k\cdot b^j)  \nonumber \\
&&+f_0\cdot (u^j\cdot d^k+u^k\cdot d^j)+h_0\cdot (\xi ^j\cdot a^k+\xi
^k\cdot a^j)  \nonumber \\
&&+h_0\cdot (u^j\cdot b^k+u^k\cdot b^j)-h_0\cdot (u^j\cdot d^k+u^k\cdot d^j)]
\label{3.3} \\
&&+\rho \cdot (g_{\overline{l}\overline{n};m}+g_{\overline{n}\overline{r}%
}\cdot g_{l;m}^r)\cdot [(b_0\cdot u^n+h_0\cdot \xi ^n)\cdot (g^{kl}\cdot
u^j\cdot u^m  \nonumber \\
&&+g^{jl}\cdot u^k\cdot u^m-g^{ml}\cdot u^j\cdot u^k)  \nonumber \\
&&+\frac 12\cdot (c_0\cdot \xi ^n+f_0\cdot u^n)\cdot [g^{kl}\cdot (u^j\cdot
\xi ^m+u^m\cdot \xi ^j)  \nonumber \\
&&+g^{jl}\cdot (u^k\cdot \xi ^m+u^m\cdot \xi ^k)-g^{lm}\cdot (u^j\cdot \xi
^k+u^k\cdot \xi ^j)]]\}-p\cdot g_i^j\text{ \thinspace \thinspace \thinspace
\thinspace \thinspace \thinspace .}  \nonumber
\end{eqnarray}

\subsection{Variational energy-momentum tensor of Euler-Lagrange}

The variational energy-momentum tensor of Euler-Lagrange $\overline{Q}_i\,^j$
could be obtained in the form 
\begin{eqnarray}
\overline{Q}_i\,^j &=&\frac{\partial p_1}{\partial u^i}\cdot u^j+b\cdot \rho
_{,i}\cdot u^j  \nonumber \\
&&+\rho \cdot g_{\overline{l}\overline{n}}\cdot [(b_0\cdot u^n+h_0\cdot \xi
^n)\cdot u^l\,_{;i}\cdot u^j+(c_0\cdot \xi ^n+f_0\cdot u^n)\cdot \xi
^l\,_{;i}\cdot u^j]  \nonumber \\
&&-g_{\overline{i}\overline{k}}\cdot \{b_0\cdot (u\rho +\rho \cdot \delta
u)\cdot u^j\cdot u^k+c_0\cdot (u\rho +\rho \cdot \delta u)\cdot \xi ^j\cdot
\xi ^k  \nonumber \\
&&+f_0\cdot (u\rho +\rho \cdot \delta u)\cdot \xi ^j\cdot u^k+h_0\cdot
(u\rho +\rho \cdot \delta u)\cdot u^j\cdot \xi ^k  \label{3.4} \\
&&+\rho \cdot [b_0\cdot (a^j\cdot u^k+a^k\cdot u^j)+c_0\cdot (b^j\cdot \xi
^k+b^k\cdot \xi ^j)  \nonumber \\
&&+f_0\cdot (b^j\cdot u^k+a^k\cdot \xi ^j)+h_0\cdot (a^j\cdot \xi
^k+b^k\cdot u^j)]  \nonumber \\
&&+\rho \cdot (g_{\overline{l}\overline{n};m}+g_{\overline{r}\overline{n}%
}\cdot g_{l;m}^r)\cdot [(b_0\cdot u^n+h_0\cdot \xi ^n)\cdot g^{kl}\cdot
u^j\cdot u^m  \nonumber \\
&&+(c_0\cdot \xi ^n+f_0\cdot u^n)\cdot g^{kl}\cdot u^m\cdot \xi
^j]\}\,\,\,\,\,\,\,\,\,\,\,.  \nonumber
\end{eqnarray}

From the explicit form of the energy-momentum tensors and the second
Noether's identity the relation 
\begin{equation}
\frac{\partial p_1}{\partial u^i}\cdot u^j=\frac{\partial p_1}{\partial
f_{N,j}}\cdot f_{N,i}\text{ \thinspace \thinspace \thinspace \thinspace .}
\label{3.5}
\end{equation}

\noindent follows.

\section{Special cases}

The general form of the Lagrangian density $\mathbf{L}$ could be specialized
for different from zero constants $p_0$, $a_0$, $b_0$, $c_0$, $f_0$, $h_0$, $%
\kappa _0$, $k_0$, $m_0$, $M_0$, and $k_1$. If only $p_0$, $a_0$ and $h_0$
are different from zero constants, then the corresponding Euler-Lagrange's
equations and energy-momentum tensors describe a fluid with points moving on
auto-parallel lines \cite{Manoff-6}. All more general cases are also worth
to be investigated. This will be the task of another paper.

\section{Conclusion}

In the present paper Lagrangian theory for fluids over $(\overline{L}_n,g)$%
-spaces is worked out on the basis of the method of Lagrangians with
covariant derivatives. A concrete Lagrangian density is proposed. The
Euler-Lagrange's equations and the energy-momentum tensors are found. They
could be used for considering the motion of fluids and their kinematic
characteristics. It is shown that the description of fluids on the basis of
the identification of their pressure with a Lagrangian invariant could
simplify many problems in the fluids mechanics. On the other side, every
classical field theory over spaces with affine connections and metrics could
be considered as a concrete Lagrangian theory of a fluid with given pressure.


\begin{thebibliography}{99}
\bibitem{Manoff-1}  S. Manoff, \textit{Invariant projections of
energy-momentum tensors for field theories in spaces with affine connection
and metric}.\textbf{\ }J. Math. Phys. \textbf{32} (1991) 3, 728-734.

\bibitem{Manoff-2}  S. Manoff, R. Lazov,\textit{\ Invariant projections and
covariant divergency of the energy-momentum tensors.} In \textit{Aspects of
Complex Analysis, Differential Geometry and Mathematical Physics}. Eds. S.
Dimiev, K. Sekigava, (World Scientific, Singapore 1999), pp. 289-314
[Extended version: E-print (1999) gr-qc/99 07 085].

\bibitem{Brown}  J. D. Brown, D. Marolf, \textit{Relativistic material
reference system. }Phys. Rev. \textbf{D 53} (1996) 4, 1835-1844.

\bibitem{Taub}  A. H. Taub, \textit{General relativistic variational
principle for perfect fluids.} Phys. Rev. \textbf{94} (1954) 6, 1468-1470.

\bibitem{Schutz-1}  B. F. Schutz Jr., \textit{Perfect fluids in General
Relativity: Velocity potentials and a variational principle.} Phys. Rev. 
\textbf{D 2} (1970) 12, 2762-2773.

\bibitem{Schutz-2}  B. F. Schutz, R. Sorkin, \textit{Variational aspect of
relativistic field theories, with application of perfect fluids.} Ann. of
Phys. \textbf{107} (1977) 1-2, 1-43.

\bibitem{Manoff-3}  S. Manoff, \textit{Spaces with contravariant and
covariant affine connections and metrics.} Physics of elementary particles
and atomic nucleus (Physics of Particles and Nuclei) [Russian Edition: 
\textbf{30} (1999) 5, 1211-1269], [English Edition: \textbf{30} (1999) 5,
527-549].

\bibitem{Manoff-4}  \_\_\_\_\_ , \textit{Lagrangian theory of tensor fields
over spaces with contravariant and covariant affine connections and metrics
and its applications to Einstein's theory of gravitation in }$\overline{V}_4$%
-\textit{spaces.} Acta Appl. Math. \textbf{55} (1999) 1, 51-125.

\bibitem{Manoff-5}  \_\_\_\_\_ , \textit{Lagrangian formalism for tensor
fields}. In \textit{Topics in Complex Analysis, Differential Geometry and
Mathematical Physics}. eds. Dimiev St., Sekigawa K. (World Sci. Publ.,
Singapore, 1997), pp. 177-218

\bibitem{Manoff-6}  \_\_\_\_\_ , \textit{Auto-parallel equation as
Euler-Lagrange's equation over spaces with affine connections and metrics.}
Gen. Rel. and Grav. \textbf{32} (2000) 8, 1559-1582.
\end{thebibliography}
\end{document}